\documentstyle[12pt]{article}
\begin{document}

\vspace*{0.5cm}
\baselineskip=0.8cm
\begin{center}
{\Large\bf Density Functional Theory of Spin-Polarized Disordered
Quantum Dots}
\vspace{2.0cm}

{\large\bf Kenji Hirose$^{1}$, Fei Zhou$^{2}$, and Ned S.
Wingreen$^{3}$}
\vspace{0.7cm}

$^{1}$ Fundamental Research Laboratories, NEC Corporation, 34
Miyukigaoka, Tsukuba, Ibaraki 305-8501, Japan\\
$^{2}$ Physics Department, Princeton University, Princeton, New
Jersey 08544\\
$^{3}$ NEC Research Institute, 4 Independence Way, Princeton, New Jersey
08540\\
\vspace{0.5cm}
\end{center}

\baselineskip=0.72cm

\vspace{1.5cm}

\begin{center}
{\large\bf Abstract}
\end{center}

\vspace{0.6cm}

Using density functional theory, we investigate 
fluctuations of the ground-state energy of spin-polarized,
disordered quantum dots in the metallic regime. 
To compare to experiment, we evaluate the distribution of 
addition energies and find a convolution of the Wigner-Dyson 
distribution, expected for noninteracting electrons,
with a narrower Gaussian distribution due to interactions. 
The third moment of the total distribution is independent
of interactions, and so is predicted to decrease by a factor 
of $(2 - 5 \pi/8)/(2 - 6/\pi) \simeq 0.405$ upon application 
of a magnetic field, which transforms from the Gaussian 
orthogonal to the Gaussian unitary ensemble.

\vspace{1.0cm}

\newpage

The interplay of disorder and electron-electron interactions
in quantum dots has recently attracted much attention.
Experiments using quantum 
dots -- small islands fabricated in a two-dimensional electron
gas \cite{Kastner} -- measure the spacings between conductance
peaks in the Coulomb blockade region. Since the peak spacings 
reflect differences between ground-state energies 
{\it for different numbers of electrons} one cannot apply
random matrix theory \cite{Mehta} to evaluate the spectrum of 
peak-spacing fluctuations. Indeed, experiments find a more 
symmetric distribution than the Wigner-Dyson form.
Experiments disagree, however, on the magnitude of the fluctuations 
\cite{Sivan,Simmel,Patel,Simmeltwo}. 
Sivan {\it et al.} \cite{Sivan}, observed fluctuations several 
times as large as the inferred mean level spacing 
$\langle\Delta_0\rangle$, and concluded that the fluctuations 
are a fixed percentage $10-15\%$ of the total charging energy 
$e^2/C$, where $C$ is the dot capacitance.
Similar results have recently been obtained by Simmel {\it et al.}
\cite{Simmeltwo} in small Si dots.
In contrast, Patel {\it et al.} \cite{Patel} found fluctuations
in GaAs dots comparable to the mean level spacing. 

Theoretical treatments also disagree regarding the magnitude
of the peak-spacing fluctuations. Sivan {\it et al.} \cite{Sivan} 
found large fluctuations scaling as $(0.10 - 0.17)e^2/C$ for a small
lattice model.
Similar results were found by  Koulakov {\it et al.} for the 
classical, strong interaction regime $r_s \gg 1$ \cite{rs}
where electrons form a Wigner lattice \cite{Koulakov}.
Blanter {\it et al.} \cite{Blanter} used the random phase 
approximation (RPA) \cite{Berkovits1} for weakly interacting 
dots and concluded that, for dimensionless conductance $g \gg 1$ 
\cite{g}, the contribution to fluctuations from interactions should 
be parametrically {\it smaller} than the mean level spacing
$\langle\Delta_0\rangle$. While the above results can be reconciled 
as applying to different regimes of $r_s$ and $g$, recent work 
employing the self-consistent Hartree-Fock equations 
\cite{Walker,Cohen,Levit} found peak-spacing fluctuations several 
times as large as $\langle\Delta_0\rangle$ even for $r_s \sim 1$ 
and $g \gg 1$ where RPA should still provide a good approximation 
\cite{Berkovits1}.

The purpose of the present article is to clarify the origin, 
magnitude, and distribution of peak-spacing fluctuations in 
spin-polarized disordered quantum dots in the regime 
$g \stackrel{\textstyle >}{\sim} 1$ and $r_s \sim 1$.
Density functional theory (DFT) provides us with accurate
ground-state energies including electron-electron interaction, 
confinement, and disorder for  realistic quantum dots. 
We find that the distribution of peak spacings is the convolution
of a Wigner-Dyson distribution, expected for noninteracting electrons,
with a {\it narrower} Gaussian distribution due to interactions. 
The width of the Gaussian is accurately given by the fluctuations
in the screened Coulomb interaction between a pair of electrons
at the Fermi energy -- a result which will also apply to unpolarized
quantum dots. The total peak spacing fluctuations are hence smaller 
than the mean level spacing $\langle\Delta_0\rangle$.
Use of an {\it unscreened} interaction between electrons, either 
direct or exchange, is found to greatly overestimate the magnitude 
of the fluctuations.  Furthermore, since interactions add a symmetric 
contribution to the distribution of peak-spacing fluctuations, 
the third moment of the total distribution is independent of 
interactions. Hence, we predict that experimental application of a 
magnetic field will reduce the third moment by a universal factor 
of 0.405, corresponding to a change from the Gaussian orthogonal 
ensemble to the Gaussian unitary ensemble.

The ground-state energies of spin-polarized, disordered quantum 
dots are obtained within density functional theory with the 
exchange-correlation part of the electron-electron interactions 
treated in the local-density approximation. Specifically, we solve 
the following Kohn-Sham equations \cite{Kohn} numerically, and 
iterate until self-consistent solutions are obtained \cite{twoways};
\begin{equation}
\left[-\frac{\hbar^2}{2m^*}\nabla^2
+\frac{e^2}{\kappa}\int\!\frac{\rho({\bf r'})}{|{\bf r-r'}|}d{\bf
r'}+\frac{\delta E_{\rm xc}[\rho,\zeta]}{\delta\rho({\bf r})}
+V_{\rm ext}({\bf r})\right]\Psi_i({\bf r})=\epsilon_i\Psi_i({\bf r}),
\label{eq:kseqs}
\end{equation}
where the density is
\begin{equation}
\rho({\bf r})=\sum_{i}^N|\Psi_i({\bf r})|^2.
\label{eq:density}
\end{equation}
Here $E_{\rm xc}[\rho,\zeta]$ is the exchange-correlation energy
functional \cite{Tanatar} with local spin polarization 
$\zeta({\bf r}) = 1$.
The summation in the density (\ref{eq:density}) 
is taken over the $N$ lowest energy  
Kohn-Sham orbitals. In previous work \cite{Hirose}, we have  shown 
that the DFT method gives very accurate ground state energies for 
{\it clean} parabolic GaAs quantum dots, in agreement with exact
calculations for up to five electrons \cite{accuracy}.
Comparison with quantum Monte Carlo calculations \cite{accuracytwo}
confirms that DFT is valid for interaction strengths up to 
$(e^2/\kappa\ell_0)/\hbar\omega_0=6\ (r_s \simeq 8)$ 
and up to $N=8$ electrons.

The external potential for our disordered dots is the sum of a 
confining parabola and multiple ``impurity" potentials each with a 
Gaussian profile:
\begin{equation}
V_{\rm ext}({\bf r})=\displaystyle{\frac{1}{2}m^*\omega_0^2r^2
+\frac{1}{2\pi\lambda^2}\sum_{i}^{N_{\rm imp}}{\gamma_i}\cdot
{\rm exp}\left(-\frac{|{\bf r}-{\bf r}_i|^2}{2\lambda^2}\right)}.
\label{eq:vext}
\end{equation}
The impurity potentials are randomly distributed with density 
$n_{\rm imp}=1.03\times10^{-3}\, {\rm nm}^{-2}$ and strength $\gamma_i$ 
uniformly distributed on $[-W/2,W/2]$ with $W = 10 \hbar^2/m^*$. 
The width of each impurity is taken as $\lambda=\ell_0/(2\sqrt{2})$, 
where $\ell_0=\sqrt{\hbar/m^*\omega_0} \simeq 19.5 {\rm nm}$. 
Here we use the effective mass for GaAs, $m^*=0.067m$, and
$\hbar\omega_0=3.0 {\rm meV}$. 
The strength of the Coulomb interaction is controlled by changing 
the dielectric constant $\kappa$, where $\kappa = 12.9$ for GaAs. 
The resulting dimensionless interaction strength is measured by
$(e^2/\kappa\ell_0)/\hbar\omega_0$ or $r_s(=1/\sqrt{\pi\rho_0}a_B^*)$
where $a_B^* = \hbar^2 \kappa/ m^* e^2$ is the effective Bohr radius,
and $\rho_0$ is the electron density at the center of the dot. From 
a scattering phase-shift analysis we find the mean free path
of electrons $l=v_F\tau\simeq 170{\rm nm}$ to be slightly larger
than the dot diameter $L = 120-160 {\rm nm}$, where the dot diameter 
increases with $r_s$. Therefore the dots are marginally in the ballistic 
regime and have a dimensionless conductance  $g=2-4$ \cite{g}. 

At low temperatures, electron hopping into a dot containing $N-1$
electrons is suppressed except when the ground-state free energy 
$E(N-1)-(N-1)\mu$ is equal to the ground state free energy for 
$N$ electrons $E(N)-N\mu$. This degeneracy condition determines 
the position of the $N$th conductance peak as a function of the 
electron chemical potential $\mu_N=E(N)-E(N-1)$, or equivalently, 
as a function of an applied gate voltage \cite{gatevoltage}. 
The increase in $\mu$ needed to put an extra electron in the dot, 
which we will refer to as the addition energy $\Delta$, is given by 
$\Delta=E(N+1)-2E(N)+E(N-1)$. From our solution of the Kohn-Sham
equations, the ground-state energy of a dot with $N$ electrons
is obtained from
\begin{equation}
E(N)=\sum_{i}\epsilon_i-\frac{e^2}{2\kappa}\int\frac{\rho({\bf
r})\rho({\bf r'})}{|{\bf r-r'}|}d{\bf r}d{\bf r'}-\int\!
\rho({\bf r})\frac{\delta E_{\rm xc}[\rho,\zeta]}
{\delta\rho({\bf r})}d{\bf r}\ +\ E_{\rm xc}.
\label{eq:gsenergy}
\end{equation}
We consider fluctuations of the addition energy for $N=10$ electrons. 
Thus for each realization of disorder we calculate 
$\Delta \equiv E(11)-2E(10)+E(9)$. 
The disorder average is taken over more than 
1,000 different impurity configurations. As a check of accuracy, 
we have confirmed that the ground-state energies obtained from DFT 
for disordered quantum dots with $N=2$ and $3$ are in good agreement 
with exact diagonalization results for $0 \leq r_s \leq 5$.

Figure 1 shows the distribution of addition energies $\Delta$ for
interacting dots with $(e^2/\kappa\ell_0)/\hbar \omega_0=2.39$ 
(right), and the distribution of $\Delta_0$ 
for noninteracting dots of the same size (left). In the inset, we show
the charge density $\rho({\bf r})$ for one realization of disorder.
The distribution of level spacings $\Delta_0$ in the noninteracting 
dots has the Wigner-Dyson form, while that in the interacting dots 
is somewhat more symmetrical. The symmetry continues to increase
with increasing interaction strength. While 
interactions considerably enhance the average addition energy 
$\langle\Delta\rangle \simeq 6.50 \langle\Delta_0\rangle$,
the fluctuation in the interacting case 
$\delta \Delta \equiv \sqrt{\langle \Delta^2\rangle 
   - \langle \Delta \rangle^2 }$ is only $\sim 13 \%$
larger than the noninteracting fluctuation
$\delta \Delta_0 \equiv \sqrt{\langle \Delta_0^2\rangle 
   - \langle \Delta_0 \rangle^2 }$.

Figure 2 shows the average addition energy $\langle\Delta\rangle$, its
rms fluctuations $\delta \Delta$, and its third moment for disordered 
dots as a function of the Coulomb interaction strength
$(e^2/\kappa\ell_0)/\hbar \omega_0$. For comparison, we have also 
plotted results for disordered, noninteracting dots of the same size, 
which we obtain as follows. First we find the effective potential for 
the clean dot, without impurities. Then we solve for the 
single-particle level energies $\epsilon_{i}^{0}$ for this effective 
potential plus the random impurity potentials in (\ref{eq:vext}).
The addition energy is simply given by
$\Delta_0=\epsilon_{N+1}^{0}-\epsilon_{N}^{0}$ 
with $N=10$. All dot sizes satisfy the relation $\delta \Delta_0=
\sqrt{4/\pi-1}\, \langle\Delta_0\rangle \simeq 0.52 
\langle\Delta_0\rangle$ predicted by random matrix theory for the 
Gaussian orthogonal ensemble \cite{Mehta}.
In Fig. 2(a), the average addition energy in the noninteracting
case is seen to decrease with increasing Coulomb interaction.
This is because the increasing Coulomb repulsion among electrons 
causes the dot to grow and hence the level spacing to shrink. 
The average addition energy in the interacting case increases 
considerably with Coulomb interaction strength, as expected from 
the classical electrostatics relation 
$\langle\Delta\rangle\simeq \displaystyle{\frac{e^2}{C}}$ where $C$ 
is the capacitance of the dot. However, Fig. 2(b) shows that the 
interactions only slightly increase the addition-energy fluctuations.
For GaAs, $r_s \simeq 2$, the enhancement is only about 10\%, in rough 
agreement with the experiment of Patel {\it et al.} \cite{Patel}.

To understand the magnitude of addition-energy fluctuations,
we use the phenomenological framework presented by Blanter, Mirlin, 
and Muzykantskii \cite{Blanter} for the regime $r_s \ll 1$, where 
RPA is valid, and show that it applies to the DFT results at least 
up to $r_s \simeq 5$. Consider first a dot containing $N-1$ electrons 
in the ground state. Addition of the $N$th electron to form the $N$ 
electron ground state requires an electron chemical potential 
$\mu_{N}$. To form instead the first excited state of $N$ electrons
requires the higher chemical potential $\mu_{N} + \Delta \epsilon$. 
For an ensemble of disordered metallic dots, $\Delta \epsilon$ 
will have Wigner-Dyson statistics, with  $\langle\Delta \epsilon\rangle$
equal to the mean noninteracting level spacing $\langle\Delta_0\rangle$,
since the lowest excitation of a Fermi liquid is a single electron
promoted across the Fermi surface. 

The addition energy $\Delta$ is the {\it increase} in chemical potential 
from $\mu_N$ required to add one more electron to the dot and thus form 
the $N+1$ electron ground state. This $(N+1)$st electron must have 
an extra energy $\Delta \epsilon$ to occupy the lowest empty level
{\it plus} an extra energy $U_{N,N+1}$ due to its Coulomb interaction 
with the $N$th electron. The total addition energy $\Delta$ will be 
approximately given by the sum of these two contributions,
\begin{equation}
\Delta \simeq \Delta \epsilon + U_{N,N+1}.
\label{eq:delta}
\end{equation}
The distribution of $\Delta \epsilon$ is given by the 
Wigner-Dyson distribution of level spacings for a noninteracting dot 
of the same size. The average interaction energy 
$\langle U_{N,N+1}\rangle$ is the capacitive charging energy $e^2/C$. 
We estimate the fluctuations in $U_{N,N+1}$ by calculating the 
screened Coulomb interaction between two electrons at the Fermi 
surface \cite{Blanter}. Specifically, we treat the screening effect 
in the Thomas-Fermi approximation as
\begin{equation}
{U}_{N,N+1}^{\,TF}=\displaystyle{e\!\int\! {\varphi}_{N}({\bf r})
\rho_{N+1}^{0}({\bf r})d{\bf r}}. 
\label{eq:TFenergy}
\end{equation}
The screened potential due to the $N$th electron in Fourier
representation is \cite{TF}
\begin{equation}
{\varphi}_{N}({\bf q})=\displaystyle{\frac{2\pi 
e}{\kappa}\frac{\rho_{N}^{0}({\bf
q})}{|{\bf q}|+q_0}}, 
\label{eq:screening}
\end{equation}
where $\rho_{N}^{0}({\bf r})=|\phi_{N}^{0}({\bf r})|^{2}$ 
is the density of the $N$th single-particle wavefunction 
$\phi_N^0({\bf r})$ of a noninteracting disordered dot. 
The Thomas-Fermi wavevector is $q_0=(2\pi e^2/\kappa)(dn/d\mu)= 1/a_B^*$. 
It is found that the fluctuation
$\delta {U}^{\,TF} = \sqrt{ \langle({U}_{N,N+1}^{\,TF})^2 \rangle
- \langle{U}_{N,N+1}^{\,TF}\rangle^2}$  
is always considerably smaller than the noninteracting level-spacing 
fluctuation $\delta \Delta_0$ up to at least $r_s \simeq 5$.
The total fluctuation estimated as $\delta {\Delta}^{TF}=\sqrt{(\delta 
\Delta_0)^2 + (\delta {U}^{\,TF})^2}$
is shown in Fig. 2(b) by crosses. We see that the fluctuations in 
the Thomas-Fermi screening model agree well with the DFT results with 
no free parameters. This supports the picture \cite{Blanter} 
that the addition-energy fluctuation arises from two quasi-particles 
above a filled Fermi sea interacting via a {\it screened} Coulomb potential. 

Within this picture, the increase of the fluctuation of $U_{N,N+1}$ 
with increasing interaction strength leads naturally to greater 
symmetry of the distribution of addition energies. Numerically, we 
find that that the distribution of $U_{N,N+1}^{\,TF}$ has a symmetric 
Gaussian form. Hence, in agreement with Eq. (\ref{eq:delta}),
we observe that the addition-energy distribution function $P(\Delta)$ 
is always extremely well described by the convolution of a Wigner-Dyson 
distribution for level spacings $\Delta \epsilon$, 
\begin{equation}
P_{\rm WD}(\Delta \epsilon)=\displaystyle{\frac{\pi}{2}
\frac{\Delta \epsilon}{\langle\Delta_0\rangle^2}\,{\rm e}
^{-\frac{\pi}{4}\frac{\Delta \epsilon^2}{\langle\Delta_0\rangle^2}}} 
\label{eq:WDdist}
\end{equation}
with a Gaussian distribution for interaction energies $U_{N,N+1}$,
\begin{equation}
P_{\rm Gauss}(U_{N,N+1})=\displaystyle{\frac{1}{\sqrt{2\pi}\delta{U}}\,{\rm e}
^{-\frac{(U_{N,N+1}-\langle\Delta\rangle+\langle\Delta_0\rangle)^2}
{2(\delta {U})^2}}}.
\label{eq:Gaussdist}
\end{equation}
The result for the distribution of addition energies is 
\begin{eqnarray}
P(\Delta)&=&\displaystyle{\int\!\! \int\! d\Delta \epsilon \,
dU_{N,N+1}\, P_{\rm WD}(\Delta \epsilon)\,
P_{\rm Gauss}(U_{N,N+1})\,\delta(\Delta \epsilon+U_{N,N+1}-\Delta)}
\nonumber \\
         &=&\displaystyle{\frac{1}{2}\sqrt{\frac{\pi}{2}}\frac{\delta
U}{\alpha \langle\Delta_0\rangle^2}{
e}^{-\frac{\pi}{4}\frac{\tilde{\Delta}^2}{\alpha
\langle\Delta_0\rangle^2}} \left\{ 
e^{-\frac{\tilde{\Delta}^2}{2\alpha(\delta
U)^2}}+\sqrt{\frac{\pi}{2\alpha}}\frac{\tilde{\Delta}}{\delta U}\left[
1+{\rm erf}(\frac{\tilde{\Delta}}{\sqrt{2\alpha}\delta U}) \right]
\right\} }.
\label{eq:convolve}
\end{eqnarray}
Here $\alpha = \pi \delta U^2/(2\langle\Delta_0\rangle^2) + 1$ and 
$\tilde{\Delta}=\Delta-\langle\Delta\rangle+\langle\Delta_0\rangle$, 
where $\langle \Delta \rangle$ is the center of the distribution
and $\delta U$ is a fitting parameter giving the width of 
the fluctuations of $U_{N,N+1}$.
In the noninteracting case, $\delta {U}=0$ 
so that $P(\Delta)=P_{\rm WD}(\Delta)$
as expected. In the other limit,
$P(\Delta)$ becomes nearly symmetric for sufficiently large $\delta{U}$. 
In Fig. 1, we show $P(\Delta)$ given by Eq. (\ref{eq:convolve})
as a dashed line. It is seen that the DFT distribution is described 
very well by (\ref{eq:convolve}) with the best fit value of 
$\delta U = 0.13 {\rm meV}$ very close to the value 
$\delta U_{N,N+1}^{\,TF}=0.10 {\rm meV}$ estimated from the 
Thomas-Fermi screened Coulomb interaction between two electrons at 
the Fermi surface.

To test whether the distribution of addition energies is well
described by the sum of noninteracting level spacings and a symmetric
distribution due to interactions, we propose to compare the third 
moment of the distribution $P(\Delta)$ with and without a magnetic 
field $B_\perp$ normal to the plane of the dot. 
Since the interaction part, coming from the screened Coulomb interaction 
in our picture, is symmetric it does not contribute to the third moment 
of $P(\Delta)$. Therefore, the ratio
$\displaystyle{\frac{\langle(\Delta-\langle\Delta\rangle)^3
\rangle_{B_\perp\neq 0}}
{\langle(\Delta-\langle\Delta\rangle)^3\rangle_{B_\perp= 0}}}$ 
should take the value $(2 - 5 \pi/8)/(2 - 6/\pi) \simeq 0.405$ 
which applies to level spacings taken from a Gaussian orthogonal 
ensemble ($B_\perp=0$) and a Gaussian unitary ensemble 
($B_\perp \neq 0$) \cite{Mehta}.
Since our results apply only to the case of spin-polarized
electrons, it is necessary to apply a large magnetic field
in the plane of the dots, or to spin polarize the nuclei 
\cite{McEuen}.  The result can also be tested numerically, {\it e.g.} 
by exact diagonalization studies as in Ref. \cite{Sivan}.

Existing diagonalization studies for spin polarized electrons on 
small lattices find addition-energy fluctuations 
$\delta \Delta \simeq 0.15 e^2/C$ \cite{Sivan}.
For comparison, the Coulomb contribution to the fluctuations 
found by DFT are much smaller $ \delta \Delta \simeq 0.03 e^2/C$ 
at $r_s \simeq 2$. This difference may be attributed to differences 
in the strength of disorder: while the dimensionless conductance in
our dots is $g = 2-4$, we estimate $g = 0.1- 0.3$ in Ref. \cite{Sivan}. 
Various theoretical estimates give fluctuations 
$\delta \Delta \propto \langle\Delta_0\rangle/g$\cite{Agam} 
or $\delta \Delta \propto \langle\Delta_0\rangle/ {\sqrt g}$
\cite{Blanter}. In either case, the discrepancy between exact 
diagonalization and DFT can be attributed to the order of magnitude
difference in the dimensionless conductance $g$ in the samples studied. 
The experiments on GaAs \cite{Sivan,Simmel,Patel} have 
$r_s \sim 1$ and $g > 1$, and thus fall more closely in the range of 
interaction strengths and dimensionless conductance treated in this
paper.

Recently, several calculations \cite{Walker,Cohen,Levit} based on the 
self-consistent Hartree-Fock (SCHF) equations have found large 
fluctuations, up to $\delta \Delta \simeq 0.2 e^2/C$, 
in the same range of $r_s$ we consider.
In one case \cite{Cohen}, the dimensionless conductance is estimated 
to be $g \gg 1$, a regime where DFT predicts fluctuations an order 
of magnitude smaller. As pointed out by Walker, Montambaux, and Gefen 
\cite{Walker}, the exchange interaction in the SCHF equations is 
unscreened. To test whether the lack of exchange screening
in the SCHF approach could be responsible for the 
discrepancy with DFT, we have calculated the unscreened exchange 
interaction between two electrons near the Fermi surface in our dots
\begin{equation}
{U}_{N,N+1}^{\,\rm exch} = { {e^2}\over{\kappa}}
\int\!\!\int\frac{\phi_N^{0*}({\bf r})\phi_N^0({\bf r'})
\phi_{N+1}^{0*}({\bf r'})\phi_{N+1}^0({\bf r})}
{|{\bf r}-{\bf r'}|}d{\bf r}d{\bf r'}.
\label{eq:exchange}
\end{equation}
In Fig. 2(b), we have plotted as open circles the fluctuations 
taken by summing the unscreened exchange interaction
(\ref{eq:exchange}) with the noninteracting level spacing \cite{direct}.
It is clear that for $r_s > 1$, the unscreened exchange interaction 
noticeably overestimates the addition-energy fluctuations. In contrast, 
density functional theory correctly accounts for screening within the
electron gas, including exchange interactions \cite{DFTscreen}. 
These results suggest that the unscreened exchange interaction
in the SCHF approach may generally lead to an overestimate of the 
addition-energy fluctuations. 

In this work, we have neglected external screening by gates or 
electrodes. This simplification should be valid as long as the 
distance to external conductors is larger than the diameter of 
the dot. In the opposite limit, it is essential to consider
external screening, but this may be done by a simple modification 
of the $1/r$ potential between electrons.

In conclusion, we have studied the electronic states of 
spin-polarized, disordered quantum dots using density functional 
theory and investigated the fluctuation of the ground-state energies. 
We have found that electron-electron interactions increase the 
fluctuation of addition energies by no more than $25\%$, up to 
$r_s\simeq 5$, even though the average addition energy is increased 
by a factor of 10. The addition energy is well approximated as the 
sum of the noninteracting level spacing and the screened Coulomb 
interaction between two electrons at the Fermi surface.
Hence the distribution of addition energies is the convolution
of a Wigner-Dyson distribution of level spacings with a Gaussian 
distribution of interaction energies. Since the latter is symmetric, 
it does not contribute to the third moment of the addition-energy 
distribution. The third moment is therefore predicted to decrease 
by a universal factor of 0.405 on application of a magnetic field 
which transforms the dot from the Gaussian orthogonal to the Gaussian 
unitary ensemble. For quantum dots having larger numbers of electrons, 
whether spin-polarized or not,
we anticipate that the decrease of the screened Coulomb 
interaction-energy fluctuations occurs as fast as the decrease of 
the level spacing and thus the present results are also applicable. 

We acknowledge I.~L.~Aleiner, B.~L.~Altshuler, R.~Berkovits, 
C.~M.~Marcus, and M.~Stopa for useful comments and suggestions.

\newpage

\noindent
{\bf Figure Captions}
\begin{itemize}

\item[Figure 1:] Distribution of addition energies $\Delta$ to
add the 11th electron for interacting dots with 
$(e^2/\kappa\ell_0)/\hbar \omega_0=2.39$ (right) and 
$\Delta_0$ for noninteracting dots of the same size (left). 
The energy binwidth is 0.05 [meV]. 
The dashed lines show the distribution function obtained from Eq. (10).
Inset: The charge density profile $\rho({\bf r})$ for $N = 10$ 
electrons with one configuration of impurities.

\item[Figure 2:] (a) Average addition energy $\langle\Delta\rangle$, (b)
fluctuation $\delta \Delta$, and (c) cube-rooted third moment 
$\sqrt[3]{ \langle(\Delta - \langle\Delta\rangle)^3 \rangle}$, 
as a function of electron-electron interaction strength
$(e^2/\kappa\ell_0)/\hbar \omega_0$. The measure of interaction strength
$r_s\,(=1/\sqrt{\pi\rho_0}\,a_B^*)$ indicated by arrows in (a) can be 
applied to the data in all panels. For each data point, the disorder 
average is taken over more than 1,000 different impurity configurations. 
At each $r_s$, the noninteracting data are taken for dots of the same 
size as the interacting dots, and the relation
$ \delta \Delta_0 \simeq  0.52 \langle\Delta_0\rangle$ 
expected for noninteracting level-spacing statistics is always
satisfied. 
Also plotted in (b) are the fluctuations due to noninteracting level 
spacings plus the {\it screened} Coulomb interaction between two 
electrons at the Fermi surface (crosses), or the {\it unscreened} 
exchange interaction between the two electrons (open circles).

\end{itemize}


\begin{thebibliography}{1000}
\bibitem{rs}
The dimensionless interaction strength $r_s$ is the
interparticle spacing in units of the effective Bohr
radius, or, equivalently, the ratio of the 
typical interaction energy between electrons
to the Fermi energy.
\bibitem{g} The dimensionless conductance is given by
$g = \hbar/(\tau_L \langle \Delta_0\rangle)$ where the $\tau_L$ is the
time an electron takes to cross the dot and 
$\langle \Delta_0\rangle$ is the mean level spacing. For dots in the
ballistic regime, such as those studied here, 
$g \sim k_F L/\pi$ where $L$ is the linear
size of the dot.
\bibitem{gatevoltage} The electron chemical potential
changes linearly with gate voltage with a coefficient that depends on
the capacitance between the dot and the gate, cf. Ref. \cite{Kastner}.
\bibitem{direct} The fluctuations in the unscreened direct 
Coulomb interaction (not plotted) are $~\sim 10\%$ larger
than the fluctuations in the unscreened exchange interaction
over the range of $r_s$ considered.
\end{thebibliography}
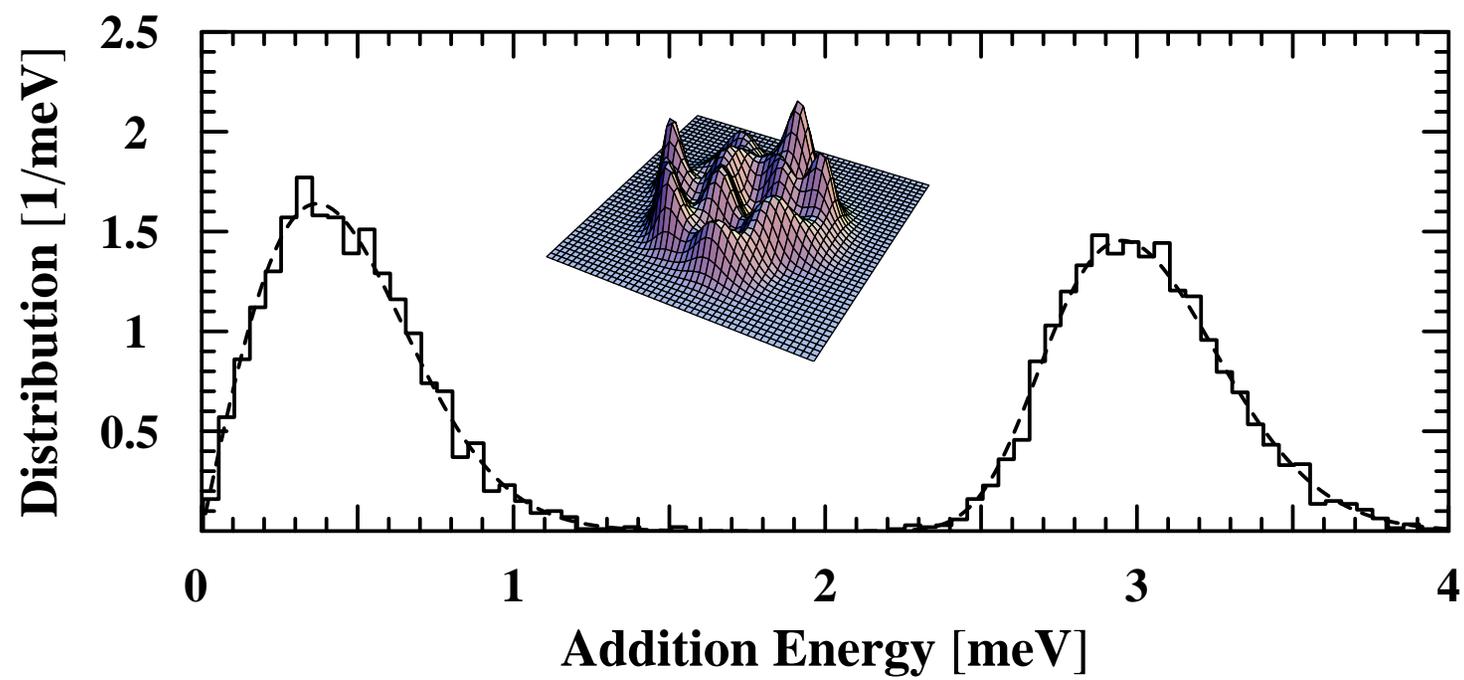
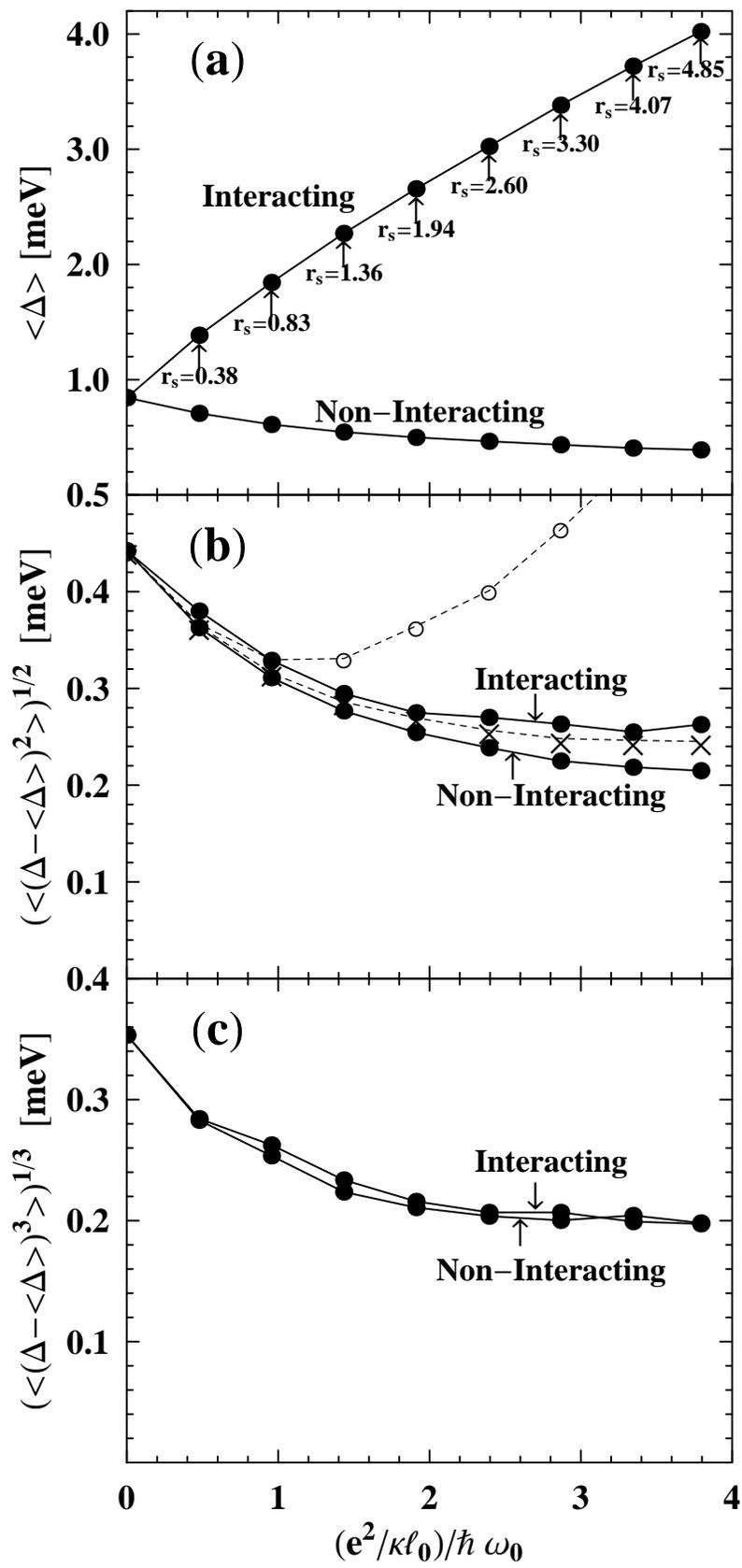
\end{document}